\documentclass[aps,preprint]{revtex4}%
\usepackage{amsfonts}
\usepackage{amsmath}
\usepackage{amssymb}
\usepackage{graphicx}%
\begin{document}
\title[Scaling Symmetry and Radiation]{Blackbody Radiation and the Scaling Symmetry of Relativistic Classical
Electron Theory with Classical Electromagnetic Zero-Point Radiation}
\author{Timothy H. Boyer}
\affiliation{Department of Physics, City College of the City University of New York, New
York, New York 10031}

\begin{abstract}
It is pointed out that relativistic classical electron theory with classical
electromagnetic zero-point radiation has a scaling symmetry which is suitable
for understanding the equilibrium behavior of classical thermal radiation at a
spectrum other than the Rayleigh-Jeans spectrum. \ In relativistic classical
electron theory, the masses of the particles are the only scale-giving
parameters associated with mechanics while the action-angle variables are
scale invariant. \ The theory thus separates the interaction of the action
variables of matter and radiation from the scale-giving parameters.
\ Classical zero-point radiation is invariant under scattering by the charged
particles of relativistic classical electron theory. \ The basic ideas of the
matter-radiation interaction are illustrated in a simple relativistic
classical electromagnetic example.

\end{abstract}
\maketitle

\section{Introduction}

Although scaling symmetry receives very little attention within classical
physics, an appreciation of this symmetry is crucial for understanding
blackbody radiation within classical theory. \ Traditional classical electron
theory with its arbitrary nonrelativistic potentials is presented as though
the scales of length, time, and energy can all be chosen independently.
\ However, this is not the scaling symmetry which appears in nature. \ Nature
has chosen a $\sigma_{ltU^{-1}}$-scaling which links together the scales of
length, time, and energy. \ The links between the scales are presented by
several fundamental constants, including the speed of light $c$ connecting
length $l$ and time $t$ ($c=l/t)$ and the elementary charge $e$ connecting
energy $U$ and length $l$($U=e^{2}/l).$ \ Here we point out that the
restrictions within $\sigma_{ltU^{-1}}$-scaling give relativistic classical
electron theory with classical zero-point radiation an additional symmetry
which gives stability to the zero-point radiation spectrum and allows the
possibility of a universal equilibrium spectrum for classical thermal
radiation. \ 

Relativistic classical electron theory with classical zero-point radiation
consists of Newton's second law for the motion of particles, all of the same
fixed charge $e$ and with various masses $m,$ under the action of the Lorentz
force due to electromagnetic fields described by Maxwell's equations, with the
homogeneous boundary condition on Maxwell's equations corresponding to random
classical electromagnetic radiation with a Lorentz-invariant spectrum,
classical electromagnetic zero-point radiation.\cite{Lorentz}\cite{Rosenfeld}%
\cite{b1975} \ The one factor setting the scale of the classical zero-point
radiation is chosen so as to provide agreement with the experimentally
observed Casimir forces and is recognizable as $(1/2)\hbar$ where $\hbar$ is
Planck's constant $h$ divided by $2\pi.$ \ Here the multiplicative constant
$\hbar$ of the classical zero-point radiation is another fundamental constant
which links energy and time. The three fundamental constants $e$, $c,$and
$\hbar$ allow the formation of one pure number with no units $e^{2}/\hbar c,$
known as the fine-structure constant. \ Classical electron theory with
classical zero-point radiation has provided calculations in agreement with
experiment for a number of phenomena which are usually regarded as lying
outside the domain of classical physics, including the Planck spectrum of
blackbody radiation, specific heats of solids, diamagnetism, Casimir forces,
van der Waals forces,\cite{delaP} and the ground state of
hydrogen.\cite{ColeH}

Although the Planck spectrum for blackbody radiation has been derived from a
number of lines of reasoning using classical physics with classical zero-point
radiation,\cite{derBb} the problem of classical radiation equilibrium has
never been completely solved within classical physics. \ Classical thermal
radiation equilibrium requires that the spectrum of thermal radiation be
stable under scattering by a "black particle." \ Previous scattering
calculations for nonrelativistic mechanical scatterers have all produced the
Rayleigh-Jeans spectrum as the equilibrium spectrum.\cite{vanV}\cite{nonlin}%
\cite{bscat}\cite{Blanco} \ Indeed, these nonrelativistic scatterers act to
transform the zero-point radiation spectrum towards the Rayleigh-Jeans
spectrum.\cite{nonlin} \ However, all of the previous scattering calculations
violate the $\sigma_{ltU^{-1}}$-scaling behavior of relativistic classical
electron theory. \ It turns out that the zero-point radiation \ spectrum is
invariant under scattering by a relativistic classical hydrogen atom.
\ Specific $\sigma_{ltU^{-1}}$-scaling behavior is needed for the universal
character of the blackbody spectrum.

The outline of this article is as follows. \ In the first part we discuss what
is meant by scaling symmetry and how it is related to the interaction of
radiation and matter. \ We note that only the Coulomb potential appears in
relativistic classical electron theory and show that it allows a separation of
the interaction of the $\sigma_{ltU^{-1}}$-scale-invariant action variables of
both matter and radiation from the $\sigma_{ltU^{-1}}$-scale-giving parameters
of mass and frequency. \ We then note that zero-point radiation is invariant
under scattering by a classical hydrogen atom. \ The second part of the
article is devoted to a simple example of a charged particle held in a
circular Coulomb orbit by a circularly-polarized plane wave. \ The example
illustrates explicitly the separation of the behavior of the $\sigma
_{ltU^{-1}}$-scale-invariant parameters from the scaling parameters in the
interaction of relativistic\ matter and radiation. \ The example also suggests
how the thermal radiation spectrum can take a universal form. \ Finally we end
with remarks on the changes in classical statistical mechanics and in
classical electron theory which are involved in our understanding of nature
within classical physics.

\section{Part I - $\sigma_{ltU^{-1}}$-Scaling Symmetry and the Interaction of
Matter and Radiation}

\subsection{Scaling Symmetry}

A set is said to be "invariant" under a scale change if the set is mapped onto
itself under the scaling operation. \ A $\sigma_{ltU^{-1}}$-scale change
simultaneously maps lengths as $l\rightarrow l^{\prime}=\sigma_{ltU^{-1}}l,$
maps times as $t\rightarrow t^{\prime}=\sigma_{ltU^{-1}}t,$ and maps energies
as $U\rightarrow U^{\prime}=U/\sigma_{ltU^{-1}}$ where $\sigma_{ltU^{-1}}$ is
chosen as some positive real number$.$\cite{scaling} \ Such a scaling
operation may be regarded as a change in the units of measurement, but
necessarily a simultaneous change of all three fundamental units. \ For an
example of $\sigma_{ltU^{-1}}$-scale invariance, consider the classical
zero-point radiation spectrum given by the set of all normal modes with a
gaussian distribution of electric field amplitudes and average energy per
normal mode satisfying the relationship $U_{\omega}=(1/2)\hbar\omega.$ \ Under
a $\sigma_{ltU^{-1}}$-scale change, the frequency $\omega$ (with units of
inverse time) of a normal mode is mapped to $\omega^{\prime}=\omega
/\sigma_{ltU^{-1}}$ while the energy $U_{\omega}$ is mapped to $U_{\omega
^{\prime}}^{\prime}=U_{\omega}/\sigma_{ltU^{-1}}.$ \ But then the functional
relationship defining zero-point radiation is unchanged since $U_{\omega
^{\prime}}^{\prime}=U_{\omega}/\sigma_{ltU^{-1}}=(1/2)\hbar\omega
/\sigma_{ltU^{-1}}=(1/2)\hbar\omega^{\prime}.$ \ Thus this distribution is
mapped onto itself, and we say that the zero-point energy spectrum is
$\sigma_{ltU^{-1}}$-scale invariant. \ On the other hand, the thermal
radiation spectrum of all the normal modes at a single temperature $T$ is not
$\sigma_{ltU^{-1}}$-scale invariant because the spectrum depends on the
temperature $T$ (with units related to energy) and is mapped onto the spectrum
depending upon temperature $T^{\prime}=T/\sigma_{ltU^{-1}}.$ \ However, the
one-parameter \textit{collection} labeled by $T$ of all thermal radiation
spectra is indeed $\sigma_{ltU^{-1}}$-scale invariant because the
\textit{collection }is mapped onto itself. \ 

An individual mass $m$ is not $\sigma_{ltU^{-1}}$-scale invariant since under
scaling $m$ (with units related to energy) is mapped to $m^{\prime}%
=m/\sigma_{ltU^{-1}}.$ \ However, the \ one-parameter \textit{collection
}labeled by\textit{ }$m$ of all masses is $\sigma_{ltU^{-1}}$-scale invariant.
\ The individual constants $e$, $c$, and $\hbar$ are all $\sigma_{ltU^{-1}}%
$-scale invariant because they have dimensions such that the factors of
$\sigma_{ltU^{-1}}$ appearing in a simultaneous scale change cancel
completely. \ We speak of a "scaling variable" or a "scaling parameter" as one
which changes under the action of a $\sigma_{ltU^{-1}}$-scale change. \ Thus
for example, the mass $m$ of a particle, the frequency $\omega$ of a normal
mode, and the temperature $T$ of a system at equilibrium are all scaling
parameters. \ On the other hand, the charge of a particle and its angular
momentum are not scaling parameters since they are $\sigma_{ltU^{-1}}$-scale invariant.

\subsection{ $\sigma_{ltU^{-1}}$-Scaling Symmetry and Adiabatic Compression}

Classical electromagnetic theory contains no fundamental length.
\ Accordingly, for pure radiation in an spherical enclosure, a $\sigma
_{ltU^{-1}}$-scale change is indistinguishable from a spherical adiabatic
compression; the normal mode frequencies, the normal mode energies, and the
volume all change in the same way.\cite{scaling} \ However, if the enclosure
contains radiation and also a mass $m$, then a change of scale is very
different from an adiabatic compression of the enclosure. \ Under a
$\sigma_{ltU^{-1}}$-scale change, the radiation and the mass will all be
mapped to new values; however, under an adiabatic compression, the scales of
radiation energy, frequency, and volume will all change while leaving the mass
$m$\ of the particle unchanged. \ The great fascination of blackbody radiation
during the nineteenth century was its universal character. \ How could the
radiation spectrum which corresponded to thermal equilibrium be independent of
the mass $m$ of a charged particle which scattered the radiation? \ How could
the \textit{form} of the equilibrium radiation spectrum in the enclosure be
unchanged after an adiabatic compression which altered the ratios of the
parameters of the radiation to the particle mass $m$ in the enclosure?

Modern physics has provided an answer to this question by changing the rules
of interaction away from classical physics and over to quantum theory.
\ However, we wish to point out that a solution within classical physics
consists in simply insisting on relativistic theory with fixed charge $e$
rather than allowing the mixtures of relativistic and nonrelativistic systems
which have appeared in previous classical analyses. \ \ Many physicists do not
seem to be aware of the no-interaction theorem of Currie, Jordan, and
Sudarshan,\cite{CJS} and the fact that special relativity imposes stringent
restrictions on the interactions between particles.\cite{restrict}
\ Relativistic classical electron theory requires that the particles interact
through electromagnetic fields. \ The particles can not interact through
potentials other than the Coulomb potential which arises from electromagnetic
fields. \ Indeed relativistic classical electron theory with classical
electromagnetic zero-point radiation has particle masses and normal mode
frequencies as essentially the only scaling parameters, and therefore the
theory has a $\sigma_{ltU^{-1}}$-scaling behavior which is different from any
theory which allows arbitrary interaction potentials. \ The $\sigma_{ltU^{-1}%
}$-scaling behavior provides the additional symmetry which decouples the
action variables from the scale-giving variables. \ This decoupling is
precisely what is wanted for a universal character of thermal radiation. \ In
the next section, we will discuss this decoupling, then in Part II we will
give a specific model example of the connection between a relativistic
scatterer and radiation. \ 

\subsection{Action-Angle Variables for Systems}

The action-angle variables $J,\theta$ of oscillating systems have dimensions
which make them invariant under $\sigma_{ltU^{-1}}$-scale changes. \ Thus, for
example, the angular momentum of a particle of mass $m$ is an action variable
having dimensions of $mass\times velocity\times length.$ \ Since mass
transforms as an inverse length while the velocity is invariant under
$\sigma_{ltU^{-1}}$-scale changes, the angular momentum is indeed
$\sigma_{ltU^{-1}}$-scale invariant.

Each of the normal modes of radiation oscillation in a cavity can be regarded
as an independent harmonic oscillator system.\cite{Power} \ When expressed in
terms of action-angle variables, the energy $U_{\omega}$ of a mode is related
to the frequency $\omega$ of the mode and the action variable $J_{\omega}$ as%
\begin{equation}
U_{\omega}=J_{\omega}\omega
\end{equation}
But then the ratio $U_{\omega}/\omega$ is given by
\begin{equation}
U_{\omega}/\omega=J_{\omega}%
\end{equation}
and is $\sigma_{ltU^{-1}}$-scale invariant, just as the action variable
$J_{\omega}$ is $\sigma_{ltU^{-1}}$-scale invariant. \ Thus for the radiation
mode, we find here that the ratio between the scale-giving parameter $\omega$
and the energy $U_{\omega}$ depends solely on the $\sigma_{ltU^{-1}}%
$-scale-invariant action variable $J_{\omega}$. \ Under an adiabatic
compression, the frequency $\omega$ of the mode will change and the energy
$U_{\omega}$ will change, but the action variable $J_{\omega}$ is an adiabatic
invariant and will not change. \ Thus a $\sigma_{ltU^{-1}}$-scale change is
the same as an adiabatic compression of the mode. \ The action variable of the
radiation is not changed in magnitude in either case. \ For this system there
is a complete decoupling of the ratio of the scale-giving parameter $\omega$
and the energy $U_{\omega}$ from anything but the action variable $J_{\omega
}.$

In relativistic classical electron theory with classical electromagnetic
zero-point radiation, the charges interact through the electromagnetic fields.
\ In the mechanical approximation which excludes radiation, the particles of
relativistic classical electron theory can be regarded as interacting through
the Coulomb potential. \ As an example of this situation, we consider a
particle of charge $e$ and mass $m$ in the Coulomb field of another particle
of charge $-e$ and very large mass, corresponding to a classical hydrogen
atom. \ The energy of the system (excluding the self-energy of the large mass)
is given by\cite{Goldstein}%
\begin{align}
U_{m}  &  =m\gamma c^{2}-e^{2}/r\nonumber\\
&  =mc^{2}\left(  1+\frac{(e^{2}/c)^{2}}{\{J_{3}-J_{2}+[J_{2}^{2}%
-(e^{2}/c)^{2}]^{1/2}\}^{2}}\right)  ^{-1}%
\end{align}
where $\gamma=(1-v^{2}/c^{2})^{-1/2},$ while$\ J_{2}$ and $J_{3}$ are the
action variables for the hydrogen system. \ If we divide through by $mc^{2}$
to obtain $U_{m}/\left(  mc^{2}\right)  ,$ then this ratio is equal to a
function of the action variables (and the\ fixed constant $e^{2}/c)$ alone,
and is not dependent on any scale-giving parameter. \ Once again we have a
decoupling of the ratio of the system's scale-giving parameter $m$ and the
energy $U$ from anything but the action variables $J_{i}$. \ Indeed, all
lengths, times, and energies for the relativistic hydrogen orbits will involve
respectively the fundamental length $e^{2}/(mc^{2})$ times a function of the
$J$'s, the fundamental time $e^{2}/(mc^{3})$ times a function of the $J$'s,
and the fundamental energy $mc^{2}$ times a function of the $J$'s. \ The
orbital speed of the mass $m$ is invariant under $\sigma_{ltU^{-1}}$-scale
change and will involve the $J$'s only. \ 

These relations can be seen easily for the restricted case of a circular orbit
where Newton's second law gives%

\begin{equation}
m\gamma\frac{v^{2}}{r}=\frac{e^{2}}{r^{2}}%
\end{equation}
while the angular momentum $J$ is%
\begin{equation}
J=m\gamma vr
\end{equation}
Then combining equations (4) and (5), the speed \ of the particle in its orbit
is%
\begin{equation}
v=\frac{e^{2}}{J}%
\end{equation}
the radius is
\begin{equation}
r=\frac{J}{m\gamma v}=\left(  \frac{e^{2}}{mc^{2}}\right)  \left(  \frac
{Jc}{e^{2}}\right)  ^{2}\left[  1-\left(  \frac{e^{2}}{Jc}\right)
^{2}\right]  ^{1/2}%
\end{equation}
the frequency is
\begin{equation}
\omega=\frac{v}{r}=v\frac{m\gamma v}{J}=\left(  \frac{mc^{3}}{e^{2}}\right)
\left(  \frac{e^{2}}{Jc}\right)  ^{3}\left[  1-\left(  \frac{e^{2}}%
{Jc}\right)  ^{2}\right]  ^{-1/2}%
\end{equation}
and the energy is%
\begin{equation}
U=mc^{2}\left[  1-\left(  \frac{e^{2}}{Jc}\right)  ^{2}\right]  ^{1/2}%
\end{equation}
In each case we see the appearance of the characteristic length $e^{2}%
/(mc^{2}),$ time $e^{2}/(mc^{3}),$ or energy $mc^{2}$ times a function of the
angular momentum $J.$

It should be emphasized that the decoupling of the $\sigma_{ltU^{-1}}%
$-scale-invariant ratios from the scaling parameters is something which occurs
only for the Coulomb potential and is not the usual situation for mechanical
systems. \ Thus, for example, the energy of the nonlinear oscillator can be
rewritten in terms of action variables as\cite{Born}
\begin{align}
H  &  =p^{2}/(2m)+m\omega_{0}^{2}x^{2}/2+\Gamma x^{3}/3\nonumber\\
&  =J\omega_{0}-\frac{5\Gamma^{2}(\omega_{0}J)^{2}}{12\omega_{0}m^{3}%
}+O(\Gamma^{3})
\end{align}
This system has three scaling parameters $m,$ $\omega_{0},$ and $\Gamma$\ with
the dimensions of $mass$, $(time)^{-1}$, and $energy\times$($length$)$^{-3}$
respectively $.$ \ The $\sigma_{ltU^{-1}}$-scale-invariant ratio $H/\omega
_{0}=J-(5\Gamma^{2}J^{2})/(12m^{3})+O(\Gamma^{3})$ is not a function\ of the
action variable $J$ alone but rather depends also upon both $m$ and $\Gamma.$

\subsection{Invariance of Zero-Point Radiation Under Scattering by a Classical
Hydrogen Atom}

Classical electromagnetic zero-point radiation consists of the radiation
normal modes at all frequencies with an energy ($1/2)\hbar\omega$ per normal
mode and random phases between the radiation modes. \ Each mode involves the
electromagnetic fields oscillating with their own initial phase, at their own
frequency $\omega,$ and with their action variable $J_{\omega}$ taking the
common value $J_{\omega}=(1/2)\hbar.$ \ Classical thermal radiation at
temperature $T>0$ has the same properties except that the action variable
$J_{\omega}$ no longer has a common value but rather varies with the frequency
$\omega,$ $J_{\omega}=F(\hbar\omega/(k_{B}T))>(1/2)\hbar$ where $F$ is some
characteristic function$.$ \ The last inequality means that each normal mode
has an energy which is larger than the zero-point energy of the mode. \ The
difference between the energy at finite temperature and the zero-point energy
gives the thermal energy in a radiation mode. \ The sum of the thermal
energies over all the modes gives the finite (for finite volume) thermal
energy in the volume. The zero-point radiation spectrum where each action
variable $J_{\omega}$ take a common value is Lorentz-invariant, $\sigma
_{ltU^{-1}}$-scale invariant, and invariant under adiabatic
compression.\cite{b1975} \ Thermal radiation at $T>0$ has a preferred Lorentz
frame where it is isotropic, and changes temperature $T$ under a
$\sigma_{ltU^{-1}}$-scale$\ $ change or under an adiabatic compression.

Classical electromagnetic radiation can not bring itself to thermal
equilibrium. \ Rather it is the interaction with matter which redistributes
the thermal radiation energy (above the zero-point energy) into the various
normal modes so as to achieve equilibrium. \ Since thermal radiation
equilibrium is determined by matter, it is clear that the symmetries of matter
will profoundly affect radiation equilibrium. \ In particular, the
$\sigma_{ltU^{-1}}$-scaling behavior of relativistic classical electron theory
with classical zero-point radiation will enforce quite different conditions
from those allowed by scatterers which have a different scaling behavior.

If the classical hydrogen atom of the previous section is placed in classical
zero-point radiation, then the radiation will interact with the orbiting
particle of charge $e$ and mass $m,$ and so the radiation will be scattered.
\ However, the distribution of values for all the action variables $J$ will be
determined independently of frequency $\omega$ or of mass $m$ since the action
variables are decoupled from the scale-giving parameters. \ The incident
zero-point radiation in free space has the action variables $J_{\omega}$ take
the same value $J_{\omega}=(1/2)\hbar$ at each frequency $\omega.$ \ Since
there is only one scale-giving parameter $m$ for the interaction between
zero-point radiation and the Coulomb system, there is no $\sigma_{ltU^{-1}}%
$-scale-invariant quantity which can be formed which involves $m$. \ Thus all
the action variables of the entire system must be functions of $\hbar$ and the
pure number $e^{2}/(\hbar c)$ which can be formed from the $\sigma_{ltU^{-1}}%
$-scale-invariant quantities $e,$ $\hbar,$ and $c.$ \ But then the spectrum of
the radiation must be $\sigma_{ltU^{-1}}$-scale-invariant, and the only
$\sigma_{ltU^{-1}}$-scale-invariant spectrum is that of zero-point radiation.
\ The hydrogen scattering system can not alter the spectrum of zero-point
radiation. \ 

The situation for a classical hydrogen atom scatterer discussed here is
totally different from the previous scattering calculations appearing in the
literature.\cite{vanV}\cite{nonlin}\cite{bscat}\cite{Blanco} \ All of the
previous scattering calculations involve non-Coulomb systems which do not have
the $\sigma_{ltU^{-1}}$-scaling symmetry of hydrogen. \ All the previous
scatterers involve several mechanical parameters, as does, for example, the
charged nonlinear oscillator given in Eq. (10) where $m,$ $\omega_{0}$, and
$\Gamma$ are all available parameters. \ Indeed, scattering calculations have
been carried out using this charged nonlinear oscillator.\cite{nonlin} \ One
finds that the scattered radiation depends explicitly on the parameters
$\Gamma$ and $\omega_{0},$ and the scattering pushes the zero-point radiation
spectrum towards the Rayleigh-Jeans spectrum. \ In addition, it should be
emphasized that all the previous scattering calculations involved systems
which were not Lorentz invariant. \ The classical hydrogen atom of
relativistic classical electron theory with classical electromagnetic
zero-point radiation is a relativistic scattering system. \ The relativistic
scatterer applied to the relativistically invariant zero-point spectrum would
be expected to produce a relativistically invariant spectrum; i.e. to leave
the zero-point spectrum invariant. \ The restrictions associated with
$\sigma_{ltU^{-1}}$-scaling behavior show that no other solution is possible.

\section{Part II - Relativistic Example of the Interaction of Matter and
Radiation Showing a Universal Radiation Spectrum}

\subsection{A Simple Example}

The ideas of the previous discussion can be illustrated by a simple example
showing the relativistic interaction of matter and radiation. \ The
calculation gives an insight into the possibilities of classical radiation
equilibrium. \ Our model starts with a charged particle $e$ of mass $m$ in
circular orbit with angular momentum $J$ in a central potential $V(r),$ taken
for convenience of calculation as $V(r)=-k/r^{n}.$ \ Since the particle is
charged, it emits radiation. \ We ask for the circularly polarized plane wave
of minimum amplitude $E_{0}$\ incident perpendicular to the orbit which will
keep the charged particle in its orbit by providing the energy lost to
radiation.. \ A spectrum of radiation amplitude $E_{0}$ versus frequency
$\omega$\ is obtained by changing the mass $m$ of the orbiting charged
particle. \ (The effects of the magnetic field of the plane wave can be
ignored. The magnetic Lorentz force can be cancelled either by supporting the
particle on a frictionless surface or by introducing two circularly polarized
plane waves propagating in opposite directions which are in phase and so have
no magnetic field at the orbit of the particle.)\cite{model}

The centripetal acceleration for the charge is provided by the force from the
potential
\begin{equation}
m\gamma\frac{v^{2}}{r}=\frac{\partial V}{\partial r}=n\frac{k}{r^{n+1}}%
\end{equation}
while the angular momentum is given by%
\begin{equation}
J=m\gamma vr
\end{equation}
Combining these two equations (11) and (12), the particle speed $v$ is given
by the solution to
\begin{equation}
\left(  \frac{v}{c}\right)  ^{2-n}\left(  1-\frac{v^{2}}{c^{2}}\right)
^{(n-1)/2}=\frac{nkm^{n-1}}{c^{2-n}J^{n}}%
\end{equation}

The power emitted by the charged particle is given by $P_{emitted}%
=(2/3)(e^{2}/c^{3})\omega^{4}\gamma^{4}r^{2}$ while the power delivered to the
charge by the incident circularly-polarized wave of electric field amplitude
$E_{0}$ (when the field is oriented parallel to the particle's velocity so as
to provide maximum power) is $P_{absorbed}=eE_{0}v.$ In steady state, the
circularly polarized plane wave must have the same frequency $\omega$ as the
orbital motion of the mass $m,$ and the power emitted must equal the power
absorbed by the particle%
\begin{equation}
\frac{2}{3}\frac{e^{2}}{c^{3}}\omega^{4}\gamma^{4}r^{2}=eE_{0}v
\end{equation}
Since the velocity $v$\ is related to the frequency $\omega$\ by $v=\omega r,$
we find
\begin{equation}
E_{0}=\frac{2}{3}\frac{e}{c^{3}}\omega^{2}v\gamma^{4}%
\end{equation}
Now the electric field $E_{0}$ has the units of $(electric$
$charge)/(length)^{2}$ so that $E_{0}/\omega^{2}$ must be $\sigma_{ltU^{-1}}%
$-scale invariant. \ Indeed from Eq. (15), we find%
\begin{equation}
\frac{E_{0}}{\omega^{2}}=\frac{2}{3}\frac{e}{c^{3}}v\gamma^{4}%
\end{equation}
where that the right-hand side depends upon the $\sigma_{ltU^{-1}}%
$-scale-invariant velocity $v$ and $\sigma_{ltU^{-1}}$-scale-invariant
constants $e$ and $c.$ \ 

In general, the $\sigma_{ltU^{-1}}$-scale-invariant ratio $E_{0}/\omega^{2}$
for the incident radiation will have a complicated dependence upon the
parameters of the mechanical system. \ As seen in Eq. (13), the speed $v$ of
the particle depends upon the quantity [$nkm^{n-1}/(c^{2-n}J^{n})],$ and
dependence upon this quantity continues into the expression (16) for
$E_{0}/\omega^{2}.$ \ Thus in general the $\sigma_{ltU^{-1}}$-scale invariant
ratio $E_{0}/\omega^{2}$ for the radiation depends upon the mechanical angular
momentum $J,$ the mechanical particle mass $m,$ and and the strength $k$ of
the mechanical potential. \ Accordingly, the radiation spectrum $E_{0}$ versus
$\omega$ obtained by varying the mass $m$ through all possible values while
holding the angular momentum $J$ fixed is not universal but rather depends
upon the choice of the potential strength $k$. \ 

We also note that for a general potential $V(r)=-k/r^{n},$ the potential
strength $k$ can not be a $\sigma_{ltU^{-1}}$-scale-invariant constant. \ The
potential strength $k=-Vr^{n}$ is transformed under a $\sigma_{ltU^{-1}}%
$-scale change as $k\rightarrow k^{\prime}=-V^{\prime}(r^{\prime}%
)^{n}=-(V/\sigma_{ltU^{-1}})(\sigma_{ltU^{-1}}r)^{n}=(\sigma_{ltU^{-1}}%
)^{n-1}k.$ \ Only for the Coulomb potential where $n=1$ $($and where
$k=e^{2})$ is this constant unchanged under a $\sigma_{ltU^{-1}}$-scale
change. \ 

Indeed, for the case $n=1$ of the Coulomb potential, the situation simplifies
enormously. \ In this case of $V(r)=-e^{2}/r$, the constant $e$ is a
$\sigma_{ltU^{-1}}$-scale-invariant constant, and the equation for the
particle orbital speed becomes $v=e^{2}/J,$ corresponding to Eq. (6) above.
\ In this case (and this case only), the particle speed does not depend upon
the particle mass $m$ for fixed particle angular momentum $J.$ \ In the
Coulomb potential, we have the $\sigma_{ltU^{-1}}$-scale-invariant ratio
$E_{0}/\omega^{2}$ of the stabilizing incident wave given by
\begin{align}
\frac{E_{0}}{\omega^{2}}  &  =\frac{2}{3}\frac{e}{c^{3}}v\gamma^{4}%
=\nonumber\\
&  =\frac{2}{3}\frac{e}{c^{3}}\frac{e^{2}}{J}\left[  1-\left(  \frac{e^{2}%
}{Jc}\right)  ^{2}\right]  ^{-2}%
\end{align}
so that this ratio depends only upon the angular momentum $J$ of the orbiting
charge and the fixed quantity $e^{2}/c.$ \ Indeed any $\sigma_{ltU^{-1}}%
$-scale-invariant quantity for the radiation is dependent entirely upon the
$\sigma_{ltU^{-1}}$-scale invariant quantity $J$\ for the mechanical system
and has no dependence upon the scaling parameter $m$. \ When we form the
spectrum $E_{0}$ versus $\omega$ of incident radiation versus frequency by
changing the mass $m$ of the orbiting charge, we find a unique radiation
spectrum $E_{0}=\omega^{2}F(J)$ where the function $F(J)$ depends only on the
mechanical $J$ at the same frequency $\omega$ (and on the quantity $e^{2}/c),$
but does not depend upon the mass $m.$ \ For fixed charge $e$, there is a
unique connection between the orbit of the particle labeled by ($J,m)$ and the
stabilizing electromagnetic radiation spectrum labeled by ($E_{0},\omega)$.
\ Indeed, if we keep $J$ at a fixed value while changing $m$, we obtain a
radiation spectrum where $F(J)$ is a constant independent of $\omega,$ which
makes the spectrum $\sigma_{ltU^{-1}}$-scale invariant. \ This spectrum is the
example's analogue of zero-point radiation. \ 

\subsection{Radiation Emission into Harmonics}

The situation of our simple example corresponds to the scattering of classical
electromagnetic radiation but not to the scattering of \textit{random}
classical radiation. \ It is the scattering of \textit{random} radiation which
is involved with thermal equilibrium. \ Nevertheless, the example gives an
idea of what is involved in the interaction of radiation and
\textit{relativistic} classical matter in the separation of the $\sigma
_{ltU^{-1}}$-scale-invariant quantities from the scaling parameters. \ Indeed,
the model also illustrates a scattering aspect of the interaction of matter
and radiation. \ The incident radiation from the circularly polarized plane
wave is partially absorbed by the orbiting charged particle; the radiation
energy is then scattered into different directions and into the harmonic
frequencies $n\omega$ of the mechanical motion. \ Indeed, the radiation
emitted per unit solid angle into the $n$th harmonic by a charge $e$ in
uniform circular motion at frequency $\omega_{0}$ is given by\cite{Jack}
\begin{equation}
\frac{dP_{n}}{d\Omega}=\frac{e^{2}\omega_{0}^{2}}{2\pi c}(n\beta)^{2}\left\{
\left[  \frac{dJ_{n}(n\beta\sin\theta)}{d(n\beta\sin\theta)}\right]
^{2}+\frac{\cot^{2}\theta}{\beta^{2}}J_{n}^{2}(n\beta\sin\theta)\right\}
\end{equation}
where $J_{n}(n\beta\sin\theta)$ is the Bessel function of order $n$ evaluated
at the argument $n\beta\sin\theta.$ \ We see that the relative radiation
emitted into the $n$th harmonic depends upon $n\beta=nv/c.$ \ However, it then
follows that for a charged particle $e$\ in a Coulomb orbit where $\beta
=e^{2}/Jc$, the relative power emitted into the harmonics is a function of the
particle action variable $J$\ alone with no dependence upon the mass $m$\ of
the orbiting particle. \ This sort of independence from the mass $m$\ is just
what is needed for stability of the zero-point radiation spectrum under
scattering by a classical hydrogen atom. \ This sort of independence does not
arise for any potential function other than the Coulomb potential.

\subsection{Connecting the Hydrogen Scatterer and the Coherent Radiation
Spectrum When $J$ is a Function of $m/T$}

In the treatment of our simple matter-radiation example thus far, we have
emphasized the case of fixed particle angular momentum $J$ independent of the
mass $m,$ and we showed that the needed incident circularly-polarized plane
wave perpendicular to the orbit of minimum amplitude corresponded to a
$\sigma_{ltU^{-1}}$-scale-invariant spectrum of electromagnetic waves. \ Now
we wish to go beyond this scale-invariant spectrum. \ Even if we choose the
particle angular momentum $J$ not as a constant (mass-independent) value but
rather choose $J$ as a function of $m/T$ for some constant $T$, we can still
obtain a unique incident radiation spectrum $E_{0}$ versus $\omega$ where the
$\sigma_{ltU^{-1}}$-scale-invariant ratio $E_{0}/\omega^{2}$\ is a functions
of $\omega/T$ where $\omega$ is the frequency of the associated wave. \ This
is exactly the sort of association which we expect at equilibrium for random
thermal radiation.

Again the crucial aspect is the decoupling of the $\sigma_{ltU^{-1}}%
$-scale-invariant quantities from the scale-carrying quantities $m$ \ for the
matter and $\omega$ for the radiation. \ The frequency $\omega$ of the
particle orbit (and also of the associated circularly-polarized plane wave in
our example) is given in Eq. (8) while the $\sigma_{ltU^{-1}}$-scale-invariant
ratio $E_{0}/\omega^{2}$ of the associated circularly polarized plane wave is
given in Eq. (17). \ The particle angular momentum $J$ is now regarded as a
function of $m/T.$ \ Then the frequency $\omega$\ of the motion given in Eq.
(8) can be divided by the temperature $T$ to give%
\begin{equation}
\frac{\omega}{T}=\left(  \frac{mc^{3}}{Te^{2}}\right)  \left(  \frac{e^{2}%
}{J(m/T)c}\right)  ^{3}\left[  1-\left(  \frac{e^{2}}{J(m/T)c}\right)
^{2}\right]  ^{-1/2}%
\end{equation}
The left-hand side of this equation is a function of $\omega/T$ while the
right-hand side is a function of $m/T$ (and $e^{2}/c).$ \ Therefore there is a
unique functional relationship between the ratio $\omega/T$ for the radiation
and the ratio $m/T$ for the matter. \ Thus the function $J(m/T)$ of the matter
can be reexpressed as a function of $\omega/T,$ and visa versa$.$ \ But then
the relationship in Eq. (17) which connects the $\sigma_{ltU^{-1}}%
$-scale-invariant ratio $E_{0}/\omega^{2}$ of the radiation to the angular
momentum $J$ of the particle can be written to give $E_{0}/\omega^{2}$ as a
function of $\omega/T.$ \ Again this is precisely the sort of connection which
we expect to hold for a classical hydrogen atom in classical thermal
radiation. \ And this connection can be made only in the case of the Coulomb
potential which will arise in relativistic classical electron theory with
classical electromagnetic zero-point radiation.

\section{Discussion}

\subsection{Classical Statistical Mechanics}

Traditional classical statistical mechanics involves equal probabilities on
phase space and a limited total energy to be distributed over the phase
space.\cite{Reif} All temperatures are treated in exactly the same fashion,
and \ there is no transition from a low-temperature to a high-temperature form
of the theory. \ Such a theory can work satisfactorily for mechanical systems
with their finite number of degrees of freedom. \ However, radiation with its
infinite number of degrees of freedom does not allow such a treatment since
the finite thermal energy will "leak out" to the divergent set of
high-frequency modes. \ Quantum mechanics changes the rules of both
electromagnetism and of statistical mechanics by introducing the idea of
quanta. \ By contrast, classical electron theory with classical
electromagnetic zero-point radiation keeps the rules of classical
electromagnetism but makes a new choice for the homogeneous boundary condition
of classical electromagnetism; this new choice invalidates the rules of
traditional classical statistical mechanics because it introduces
temperature-independent fluctuations which are present even at absolute zero.
\ \ The presence of zero-point fluctuations, which are distinct from
temperature-dependent fluctuations, means that the theory now indeed involves
a transition from a low-temperature to a high-temperature form.

In a classical theory of thermal equilibrium which includes zero-point
radiation, each allowed system must have one parameter which indicates where
the system is located in the collection of systems between zero-point energy
and high-temperature energy. \ For a radiation normal mode of frequency
$\omega$, the ratio $\omega/T$\ for the normal mode gives this location. The
action variable $J_{\omega}$ for the mode is a function of $\omega/T.$\ Thus
if there is thermal radiation at temperature $T,$ then the ratio $\omega/T$
determines the energy $U_{\omega}=$ $J_{\omega}\omega$ of the normal mode of
frequency $\omega.$ \ If $\omega/T<<1,$ the mode has its zero-point energy
value $U_{\omega}=(1/2)\hbar\omega,$ and if $\omega/T>>1,$ then the mode has
its high-temperature energy $U_{\omega}=k_{B}T.$ \ Similarly, any charged
particle of mass $m$ in a Coulomb potential $V(r)=-e^{2}/r$ must have a
zero-point energy and a thermal energy above the zero-point value. \ The ratio
$m/T$ determines the location of the particle system along the continuum from
zero-point to high-temperature value. \ 

We note that nonrelativistic classical mechanical systems which depend upon
several continuous parameters (such as a nonlinear oscillator of mass $m$,
frequency $\omega,$ and nonlinear parameter $\Gamma)$ expose ambiguous
behavior in locating the system along the low-temperature versus
high-temperature continuum. \ However, for \textit{relativistic} classical
electron theory with classical zero-point radiation, these multiparameter
systems are not possible, and the scaling situation is enormously simplified,
as is shown in the present article.

\subsection{Classical Electron Theory}

Traditional classical electron theory was introduced by H. A. Lorentz \ during
the last quarter of the nineteenth century.\cite{Lorentz} \ The theory
consisted of massive particles in nonrelativistic potentials which interacted
with electromagnetic radiation through their fixed charge $e$. \ Lorentz
assumed explicitly\cite{boundary} that the homogeneous boundary condition on
Maxwell's equations excluded fundamental electromagnetic radiation; rather all
radiation arose from the acceleration of charged particles at a finite time.
\ Traditional classical electron theory was able to account for a number of
observed phenomena, including optical dispersion, Faraday rotation, and
aspects of the normal Zeman effect.\cite{Lorentz}\cite{Rosenfeld} \ However,
as is now reported in all the text books,\cite{Eisberg} traditional classical
electron theory was not able to account for the observed Planck spectrum of
blackbody radiation nor to solve the problem of collapse for Rutherford's
atomic model. \ In the twentieth century, the idea of zero-point energy was
introduced by Planck and extended to radiation by Nernst. \ However, it was
not until the 1960s, beginning with Marshall's careful and extensive
work,\cite{Marshall} that classical electron theory with classical
electromagnetic zero-point radiation (under the title "stochastic
electrodynamics") was shown to account for some phenomena which had previously
been regarded as the exclusive domain of quantum physics, such as Casimir
forces, van der Waals forces, diamagnetism, specific heats of
solids,\cite{delaP} and even the ground state of hydrogen.\cite{ColeH} \ The
Planck spectrum of blackbody radiation held a paradoxical position. \ On the
one hand, there were a number of derivations of the Planck spectrum from
classical physics including zero-point radiation, but there were also several
scattering calculations which suggested again that classical physics led
inevitably to the divergent Rayleigh-Jeans spectrum for thermal radiation.
\ The historical situation is probably best seen in a review\cite{b1975} of
classical electron theory with classical electromagnetic zero-point radiation
written in 1975. \ In this review, one finds that the one change made away
from Lorentz's classical electron theory is the introduction of classical
electromagnetic zero-point radiation. \ There is no appreciation that the use
of nonrelativistic mechanical systems in connection with Maxwell's
relativistic electromagnetic theory is fundamentally inconsistent because the
combination satisfies neither Galilean nor relativistic invariance. \ There is
no awareness of the extreme restrictions which special relativity places on
allowed interactions. \ The theory of 1975 which incorporates nonrelativistic
potentials is valid in at most one inertial frame where the nonrelativistic
approximation for the particle system is appropriate. \ However, the analysis
given in the present work shows that stability of the thermal radiation
spectrum within classical physics can be achieved only within a fully
relativistic theory. \ In the present discussion, we have pointed out that
\textit{relativistic} classical electron theory with classical electromagnetic
zero-point radiation has the $\sigma_{ltU^{-1}}$-scaling behavior which will
leave the classical zero-point radiation spectrum invariant under scattering
by a classical hydrogen atom, and which will give the possibility of classical
thermal radiation at a spectrum with finite thermal energy.

\end{document}